\documentclass[12pt]{article}
\usepackage{graphics} \usepackage{graphicx} \usepackage{cite}

\newcommand{\bcdot}{\mbox{\boldmath$\cdot$}}
\newcommand{\btimes}{\mbox{\boldmath$\times$}}

\begin{document}
 \title{Electromagnetic Power Radiated by an Accelerating Point Charge}
\author{Jerrold Franklin\footnote{Internet address: Jerry.F@TEMPLE.EDU}\\
Department of Physics\\ Temple University, Philadelphia, PA 19122}
  \date{}
   \maketitle

\begin{abstract} 
  We derive the electromagnetic power radiated by an  accelerating point charge,
  with the acceleration and velocity in the result being taken at the present time in the motion of the accelerating charge.
   This contrasts with the usual derivation  which calculates the power radiated through the surface of a large sphere,  and gives the radiated power in terms of the acceleration and velocity at an arbitrary retarded time.
\end{abstract}

\section{Introduction} 

  Larmor's formula,\footnote{We are using Gaussian
units with $c=1$.}
  \begin{eqnarray} 
\frac{dW_{\rm rad}}{dt}
&=&\frac{2}{3}q^2a^2, 
   \label{dwdt1}
\end{eqnarray} 
     for the power radiated by an accelerating point charge,
was first derived over 100 years ago by  Joseph Larmor\cite{jl}.
The non-relativistic Larmor formula was extended for large velocities to the relativistic
Li\'enard formula,
\begin{equation}
\frac{dW_{\rm rad}}{dt}=\frac{2}{3}q^2\gamma^6[a^2
-({\bf v\btimes a})^2],
\label{prel2}
\end{equation}
by Alfred-Marie Liénard\cite{li}.
Recent derivations\footnote{See, for instance, Chapter 14 of\cite{jdj}
or Chapter 11 of\cite{dg}.}
have generally been based on the  rate, 
 \begin{eqnarray}
\frac{dW_{\rm rad}}{dt}&=&\frac{1}{4\pi}\oint_S{\bf dS\bcdot[E(r},t)\btimes{\bf B(r},t)]\nonumber\\
&=&\frac{1}{4\pi}\oint {\bf{\hat r}\bcdot[E(r},t)\btimes{\bf B(r},t)]R^2_{\rm rad}d\Omega,
   \label{Prad}
   \end{eqnarray} 
   at which radiated energy passes through a spherical surface at a large radius, 
   $R_{\rm rad}$, from the radiating particle, 
   
    However, a major problem arises if  Eq.~(\ref{Prad})  is used to calculate the electromagnetic power. 
The fields in the surface integral are to be evaluated at the present time and 
 the point $\bf r$ at which the fields are observed, but the fields for an accelerating particle are given in terms of variables given at the retarded time by the  L\'ienard-Wiechert field equations.
 
 Thus the Li\'enard formula of Eq.~(\ref{prel2}) (and the non-relativistic Larmor formula) would be given in terms of the acceleration and velocity at the retarded time, $t_r$, and not at the present time, $t$.
 The retarded acceleration and velocity would depend on the arbitrary radius picked for evaluating Eq.~(\ref{Prad}).
  That means that the acceleration and velocity appearing in Eq.~(\ref{prel2}) could be any acceleration and velocity from the past motion of the accelerating particle. That makes Eq.~(\ref{prel2}) almost useless, since the acceleration and velocity would be arbitrary, depending on what was chosen as the radius of observation, $R_{\rm rad}$, of the radiation.
  
Another problem arises when the retarded variables in the Larmor formula (as derived)  are used {\it as if} they 
were given at the present time to suggest derivations of the Abraham-Lorentz radiation reaction 
force\cite{jdj,dg}\footnote{Also see \cite{w}, which also uses the Larmor formula for a derivation of the 
Abraham-Lorentz force, and includes some earlier references.}.     

 We resolve these problems in the next section by a new derivation of Eq.~(\ref{prel2}) that gives
  the emission of electromagnetic power at the present time   in terms of all variables at the present time.
  
   \section{Electromagnetic Power Emitted\\by an Accelerating Point Charge}
  The electric and magnetic fields appearing in Eq.~(\ref{Prad})  are given 
     by the  Li\'enard-Wiechert field equations,
\begin{eqnarray}
{\bf E(r},t)&=&\left\{\frac{q({\bf\hat r}_r-{\bf v}_r)}
{r_r^2\gamma_r^2(1-{\bf\hat r}_r\bcdot {\bf v}_r)^3}\right\}
+\left\{\frac{{\bf\hat r}_r\btimes[({\bf\hat r}_r-{\bf v}_r)\btimes{\bf a}_r]}
 {r_r(1-{\bf\hat r}_r\bcdot{\bf v}_r)^3}\right\},
\label{er1}\\ 
 {\bf B(r},t)&=&{\bf{\hat r}}_r\btimes{\bf E(r},t), \label{br1}
\end{eqnarray}
 where the variables, ${\bf r}_r,{\bf  v}_r,\gamma_r=1/\sqrt{1-v_r^2}$,
 and ${\bf a}_r$ are all  evaluated at the
retarded time, 
\begin{equation} 
t_r=t-r_r.
\label{tr}
 \end{equation}
  The radius vector, ${\bf r}_r $, is the distance from the charged particle's position at
the retarded time to the point of observation of the electromagnetic
fields at the present time.  

The retarded radius,  ${\bf r}_r $,  is related to the radius,  $\bf r$, which is directed from the present position of the accelerating charge to the point of observation, by
\begin{equation}
{\bf r}_r{\bf -r=\langle v\rangle}(t_r-t)=-r_r{\bf \langle v\rangle},
\label{rr}
 \end{equation}
where ${\bf \langle v\rangle}$  is the average velocity in the interval from $t_r$ to $t$.
This means that, using Eq.~(\ref{Prad}) at a radius, $R_{\rm rad}$, the radiated power would depend not only on $R_{\rm rad}$, but also on the average velocity in the past motion of the accelerating charge.
In our derivation, we will take the limit $R_{\rm rad}\rightarrow 0$, which means that the acceleration and all the variables in Eq.~(\ref{prel2}) will be taken at the present time.

   We consider a point charge $q$ at a position ${\bf r}(t)$ with a velocity ${\bf v}(t)$ and 
   acceleration ${\bf a}(t)$.
   We make a Lorentz transformation to the rest frame of the point charge where $\bf v'=0$ and
   \begin{eqnarray}
{\bf a'_\parallel}&=&{\bf a_\parallel}\gamma^3,
\label{accl}\\
{\bf a'_\perp}&=& {\bf a_\perp}\gamma^2.
\label{acct}
\end{eqnarray}
$\bf a'_\parallel$ is the rest frame acceleration parallel to $\bf v$, and 
$\bf a'_\perp$ is the rest frame acceleration perpendicular to $\bf v$.
   
 We now evaluate the rest frame surface integral 
   \begin{eqnarray}
\frac{dW'_{\rm rad}}{dt'}&=&
\frac{1}{4\pi}\oint{\bf{\hat r}'\bcdot[E'(r'},t')\btimes{\bf B'(r'},t')]R'^2_{\rm rad}d\Omega',
   \label{Prad'}
   \end{eqnarray} 
   not at a large radius, but in the limit $R'_{\rm rad}\rightarrow 0$.
   In this limit, $t'_r\rightarrow t'$ and ${\bf a'}_r\rightarrow{\bf a'}$,
   so the electric field is given by
\begin{eqnarray}
{\bf E'(r'},t')&=&\frac{q{\bf\hat r'}}{r'^2}+\frac{[{\bf{\hat r'}({\hat r'}\bcdot a')-a'}]}{r'},
\label{er'}
\end{eqnarray}
with all variables evaluated at the present time in the rest frame.

Then, the surface integral in Eq.~(\ref{Prad'}) for the radiated power reduces to
\begin{eqnarray}
\frac{dW'_{\rm rad}}{dt'}
&=&\frac{1}{4\pi}\oint{\bf\hat r'}\bcdot[{\bf a'\btimes({\hat r}\btimes a')}]d\Omega'\nonumber\\
&=&\frac{1}{4\pi}\oint[a'^2-{\bf({\hat r}\bcdot a')^2}]d\Omega'\nonumber\\
&=&\frac{2}{3}a'^2.
   \label{Prad''}
   \end{eqnarray}
   The radiated power can be put back in terms of the original acceleration, using
   Eqs.~(\ref{accl}) and (\ref{acct}) to give
   \begin{eqnarray}
\frac{dW'_{\rm rad}}{dt'}
&=&\frac{2}{3}(a^2_\parallel\gamma^6+a^2_\perp\gamma^4)\nonumber\\
&=&\frac{2}{3}\gamma^6[a^2-({\bf v\btimes a)}^2].
   \label{Prad2}
   \end{eqnarray}
   
  The variables in Eq.~(\ref{Prad2}) are in the original moving frame, but the rate of energy emission on the left hand side of the equation is still in the rest frame.
However, the right-hand side of Eq.~(\ref{Prad2}) has been shown to be a Lorentz invariant\footnote{See, for instance, page 666 of \cite{jdj}.}, so Eq.~(\ref{Prad2}) can be Lorentz transformed to the moving frame, finally giving
  \begin{eqnarray}
\frac{dW_{\rm rad}}{dt}
&=&\frac{2}{3}\gamma^6[a^2-({\bf v\btimes a)}^2].
   \label{Prad3}
   \end{eqnarray}
   
   This result has the same form as L\'ienard's relativistic extension of Larmor's formula, but is given here 
   with all variables at the present time, and not an arbitrary retarded time
   Thus, there is no uncertainty in the radiated power given by Eq.~(\ref{Prad3}).

\section{Conclusion}

Our main conclusion is that the power radiated by an accelerating point charge
 is given by
\begin{eqnarray}
 \frac{dW_{\rm rad}}{dt}
  &=&\frac{2}{3}\gamma^6[a^2-({\bf v\btimes a)}^2].
   \label{dwdtc2}
\end{eqnarray}
 This result has the same form as Li\'enard's relativistic extension of Larmor's formula, but is given here 
   with all variables at the present time, and not at an arbitrary retarded time, with abitrary acceleration and velocity.

\end{document}